\documentclass[twocolumn,showpacs,preprintnumbers,amsmath,amssymb,nofootinbib,scrartcl]{revtex4}
\input psfig.sty
\usepackage{epsf}
\usepackage{graphicx}  
\usepackage{dcolumn}   
\usepackage{bm}        

\newcommand{\be}{\begin{equation}}
\newcommand{\en}{\end{equation}}
\newcommand{\bea}{\begin{eqnarray}}
\newcommand{\ena}{\end{eqnarray}}

\begin{document}

\title{Probing interaction in the dark sector}

\author{P. C. Ferreira$^1${\footnote{pferreira@dfte.ufrn.br}}}

\author{J. C. Carvalho${^{1,2}}${\footnote{carvalho@dfte.ufrn.br}}}

\author{J. S. Alcaniz${^2}${\footnote{alcaniz@on.br}}}

\affiliation{$^1$Departamento de F\'{\i}sica, Universidade Federal do Rio G. do Norte,  59072-970 Natal - RN, Brasil}

\affiliation{$^2$Observat\'orio Nacional, 20921-400 Rio de Janeiro - RJ, Brasil}

\date{\today}


\begin{abstract}

A phenomenological attempt at alleviating the so-called coincidence problem is {to allow} the dark matter and dark energy to interact. By assuming a coupled 
quintessence scenario characterized by an interaction parameter $\epsilon$, we investigate the precision in the measurements of the expansion rate $H(z)$ required by future experiments 
in order to detect a possible deviation from the standard $\Lambda$CDM model ($\epsilon = 0$).  We perform our analyses at two levels, namely: through Monte Carlo simulations based on $\epsilon$CDM models, in which $H(z)$ samples with different {accuracies} are generated and through an analytic method that calculates the error propagation of $\epsilon$ as a function of the error in $H(z)$. We show that our analytical approach traces simulations accurately and find that to detect an interaction {using $H(z)$ data only, these must reach an accuracy better than 1\%.}

\end{abstract}

\pacs{98.80.-k, 95.35.+d, 95.36.+x}

\maketitle


\section{Introduction}


There is nowadays significant observational evidence that the Universe is currently undergoing accelerated expansion. However, although fundamental to our understanding of the Universe, several important issues about the mechanism behind cosmic acceleration as well as its role in the cosmic dynamics remain unanswered~(see, e.g., \cite{review} for recent reviews). Among these open questions, the possibility of a non-minimal coupling between the two major energy components in the Universe, i.e. dark matter and dark energy, has been widely investigated in the current literature~\cite{ozer, wang-meng, alcaniz-lima, ernandes}.

Interacting dark matter/dark energy models violate adiabaticity and constitute a phenomenological attempt at alleviating the so-called coincidence problem (see, e.g.,~\cite{cp}). In general, these models are characterized by a dilution of the dark matter density $\rho_{m}$ which is modified  with respect to the usual $a^{-3}$ scaling, i.e.,~\cite{wang-meng,alcaniz-lima}
\begin{equation} \label{dm}
\rho_{m} \propto a^{-3 + \epsilon}\;,
\end{equation}
where $\epsilon$ can take positive and negative values depending on if the transfer of energy is from dark energy to dark matter or vice versa, respectively. 

By introducing the above result into the balance equation for the dark matter particles and dark energy, we find~\cite{ernandes}
\begin{equation} \label{rhogeral}
\rho_{\phi} ={\rho}_{\phi 0}{a}^{-3(1+w)} + \frac{\epsilon\rho_{m0}}{3|w|-\epsilon}{a}^{-3+\epsilon},
\end{equation}
where $\rho_{\phi}$ stands for the dark energy density and we have assumed the equation-of-state parameter $w = p_{\phi}/\rho_{\phi}$ relating $\rho_{\phi}$ and the dark energy pressure $p_{\phi}$.

From the observational point of view, current background tests show that the interaction parameter $\epsilon$ takes values close to zero \cite{ernandes}, which makes this class of models indistinguishable from usual $\phi$CDM scenarios ($\epsilon = 0$). In principle, this degeneracy may be broken by studying the growth of structure in these models, which is modified by the attenuated matter density evolution (Eq. \ref{dm}), and a possible distinction between coupling and uncoupled models may be verified with the upcoming data from large redshift surveys \cite{surveys}. 

In what concerns the background tests, it is worth mentioning that due to the multiple integrals that relate cosmological parameters to cosmological distances, direct determinations of the expansion rate place the tightest constraints on the dark energy equation of state $w$ by reducing the so-called smearing effect~\cite{steinhardt}. In this paper, we investigate how measurements of the expansion rate $H(z)$ may provide constraints on the interacting parameter $\epsilon$. 
To perform such analysis, we first use an analytical approach relating the accuracy of $H(z)$ measurements to that of the interacting parameter $\epsilon$. Then\textbf{,} we generate $H(z)$ samples from Monte Carlo simulations  with different accuracy and values of the interaction scale $\epsilon$ based on the matter evolution law (\ref{dm}). By considering values of $\epsilon$ that are consistent with current observations ($\epsilon \lesssim 0.1$), we show that a possible distinction between interacting and non-interacting models may be achieved only with an accuracy in $H(z)$ data better than 1\%.

\begin{figure*}[htbp] 
  	\centering
  		\includegraphics[scale=0.255]{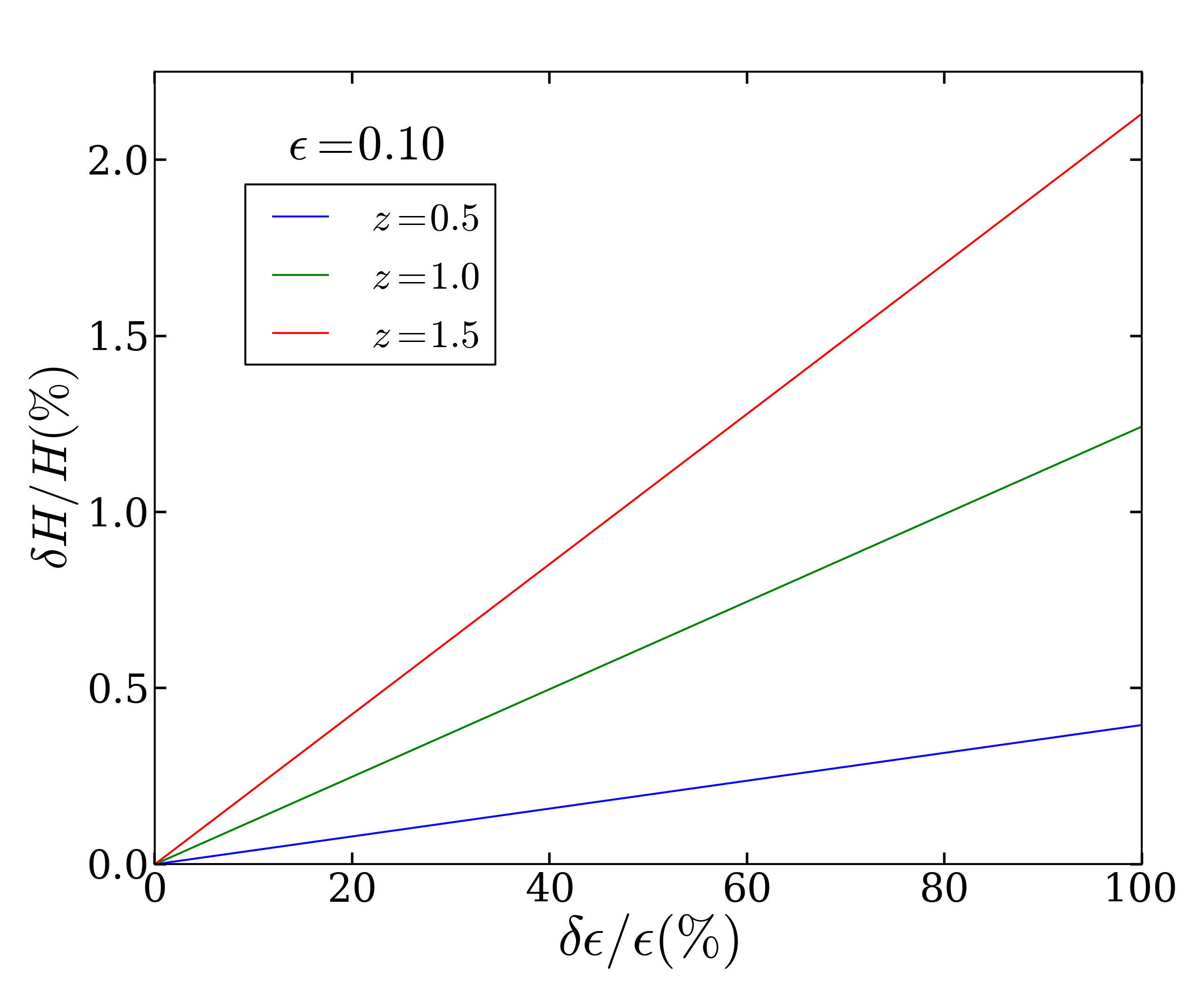}
  		\includegraphics[scale=0.255]{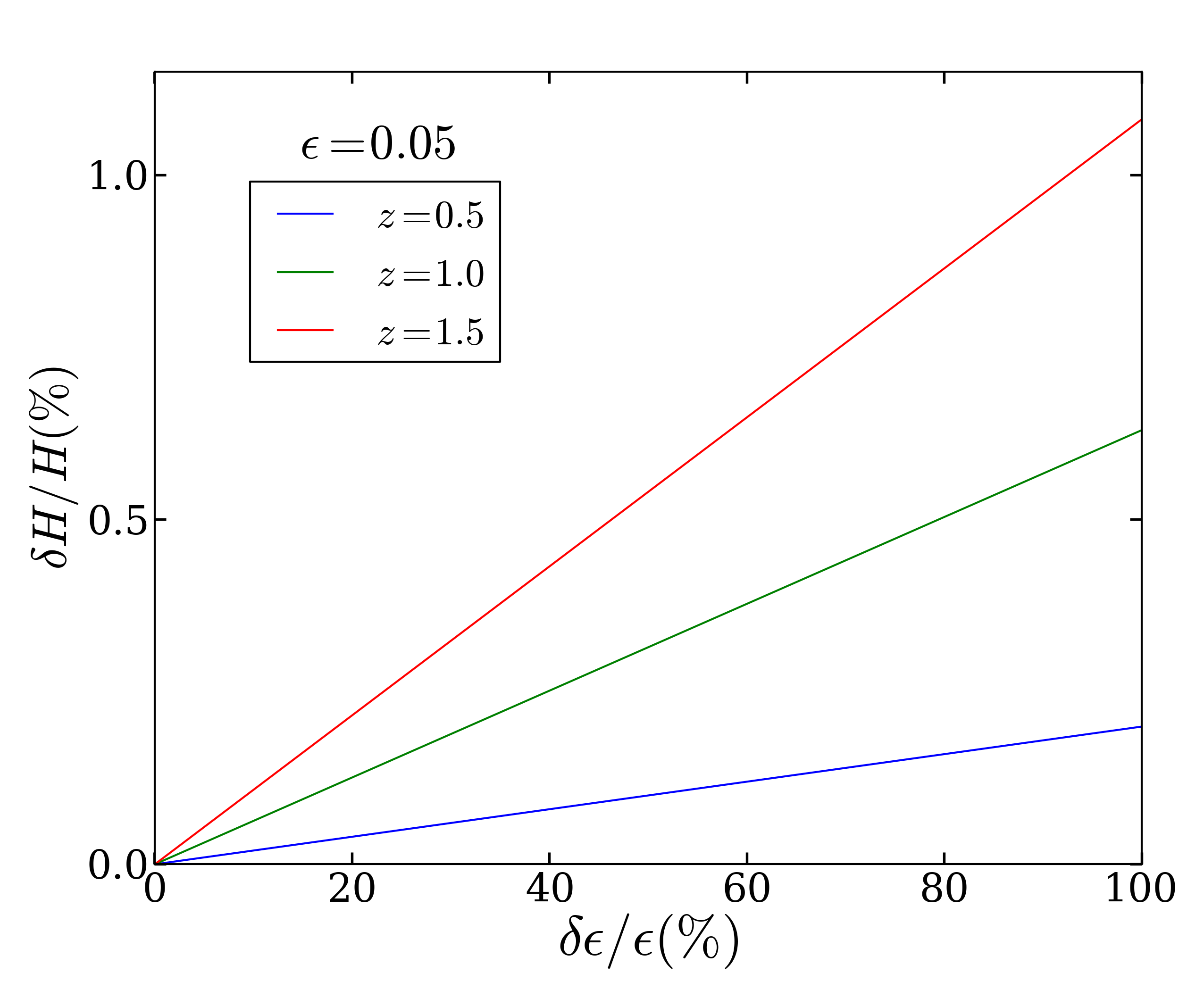}
  		\includegraphics[scale=0.255]{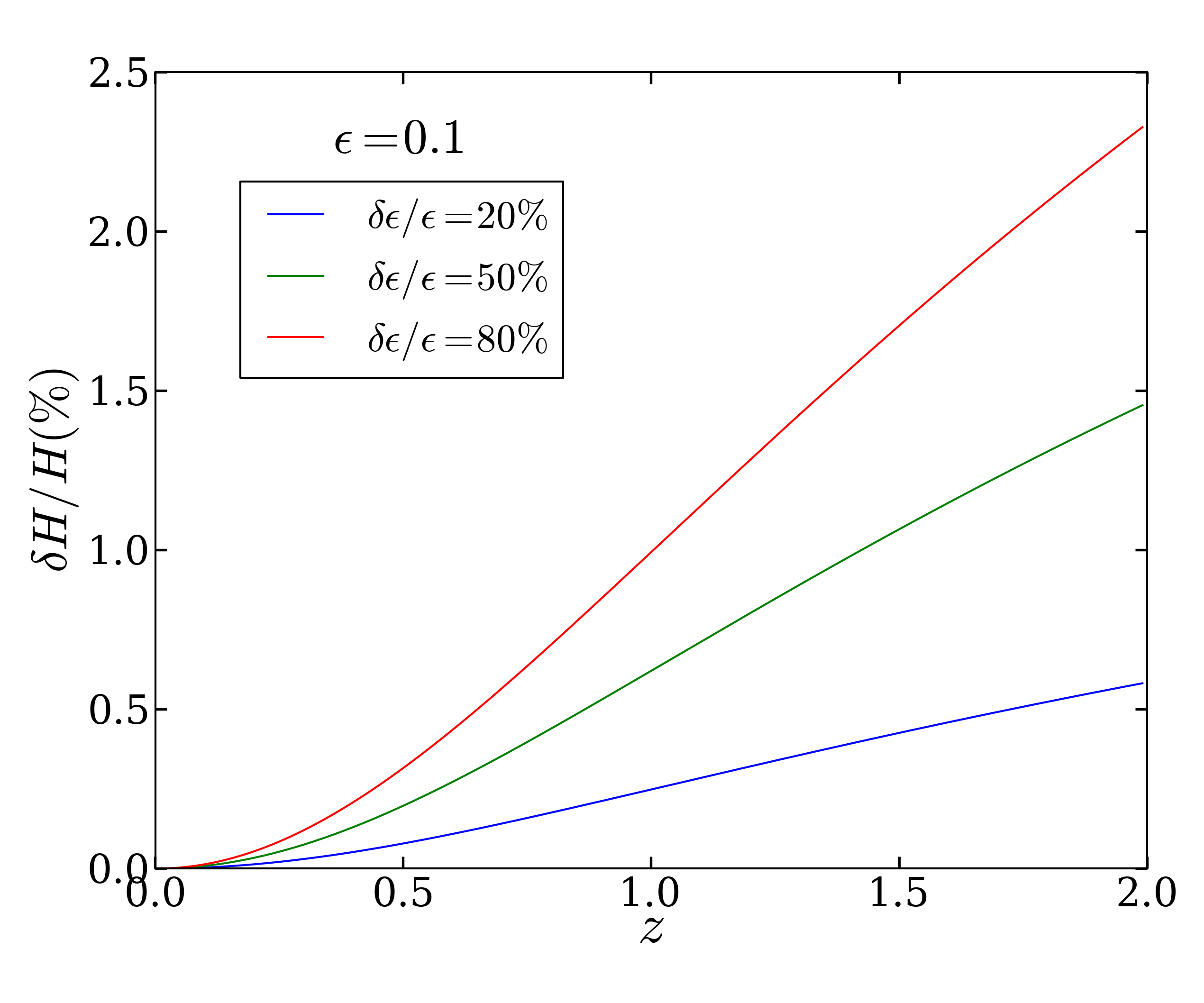}
  	\caption{Analytic computation of the relative error in $H(z)$ as a function of the relative error in $\epsilon$ for $z=0.5$, 1.0 and 1.5. The left and middle panels show the case $\epsilon = 0.1$ and $\epsilon = 0.05$, respectively. The redshift dependence of $\frac{\delta H}{H}$ for values of $\epsilon = 0.1$ and $\frac{\delta \epsilon}{\epsilon} = 20\%$, $50\%$ and $80\%$ is shown in the right panel.}
  	\label{fig:analytic}
  \end{figure*}

\section{Analytical Constraints from H(z)}

We restrict the present analysis to coupled quintessence models with $w = -1$, which is mathematically equivalent to the so-called dynamical $\Lambda$ scenarios. In this case, Eq. (\ref{rhogeral}) reduces to
\begin{equation}\label{decayv}
{\rho_{\Lambda}} =  {\tilde\rho_{\Lambda0}} + \frac{\epsilon \rho_{m0}}{3 -
\epsilon}a^{-3 + \epsilon},
\end{equation}
where the subscript 0 denotes present-day quantities. As discussed in Ref.~\cite{ernandes}, such a evolution law seems to be the most general one, having many of the previous phenomenological attempts as a particular case (see, e.g., Table I of \cite{overduin}). 

The Friedmann equation for this kind of models is given by
\begin{equation}
{H(z)} = H_0\left[\Omega_{b0}x^3 + \frac{3\Omega_{m0}}{3 - \epsilon}x^{3-\epsilon} + \tilde{\Omega}_{\Lambda0}\right]^{1/2} ,
\label{eqH}
\end{equation}
where the baryon, CDM and dark energy densities are given in units of the present critical density, $x = 1 + z$ and $\tilde{\Omega}_{\Lambda0} = \Omega_{\Lambda0} - \frac{\epsilon \Omega_{m0}}{3 - \epsilon} $. As written above, Eq. (\ref{eqH}) expresses the fact that we are considering a non-minimal coupling between the dark energy component and CDM particles only. 
We refer the reader to Ref.~\cite{freese} for a discussion about bounds on a possible interaction between dark energy and conventional matter from local gravity experiments and from primordial nucleosynthesis.

From the above equation, we calculate the relative error in the expansion rate as a function of the relative error in the interaction parameter $\epsilon$ from $\delta H^2 = \left( \frac{\partial H}{\partial \epsilon } \right)^2 \delta \epsilon^2$. This is shown in Fig. \ref{fig:analytic} for values of $\epsilon = 0.1$ (left panel) for $z = 0.5$, 1.0 and 1.5. Note that a 0.5 (1)\% error in $H$ at $z = 1.0$ results in a  40 (80)\% uncertainty in $\epsilon$, but the precision varies with redshift (right panel). For the sake of comparison, we show in the middle panel of Fig. \ref{fig:analytic} the analysis for $\epsilon = 0.05$. As expected, the smaller the value of the interacting parameter the better the accuracy in $H(z)$ measurements required to distinguish between this kind of scenario and non-interacting models with $\epsilon = 0$. In the right panel of Fig. 1 we show the dependence of $\delta H/H$ with redshift for the case $\epsilon=0.1$. For all the analysis performed in this paper, we 
consider a flat universe with $H_0 = 73$ $\rm{km.s^{-1}.Mpc^{-1}}$, $\Omega_{b0} = 0.04$ and $\Omega_{m0} = 0.23$~\cite{wmap7}.

The above analytical approach poses difficulty when one tries to compare these predictions with data. The reason is that it depends on a specific value of the redshift whereas to fit the data (real or simulated) we use information of various data points spread over a redshift interval. In order to take that into account we averaged $\delta H$ over a redshift interval by using
\begin{equation}
  \overline{\delta H} = \left[ \frac{1}{z_f - z_i} \int_{z_i}^{z_f} \left| \frac{\partial H}{\partial \epsilon} \right| dz \right] \delta \epsilon \;,
  \label{eqdeltaH}
\end{equation}
where $z_i = 0.1$ and $z_f = 1.5$.

\begin{figure*}[htbp] 
  	\centering
  		\includegraphics[scale=0.34]{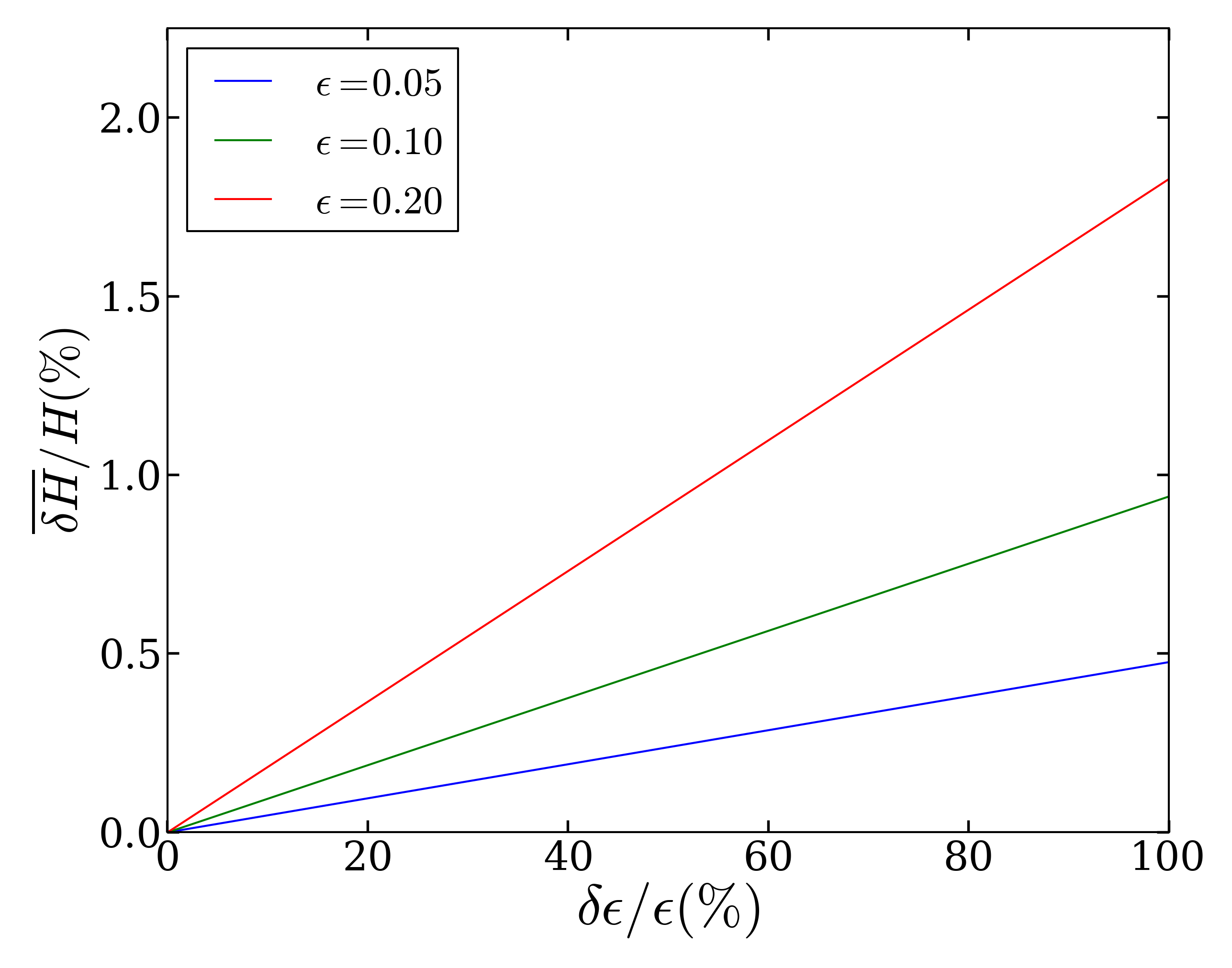}
  		\includegraphics[scale=0.34]{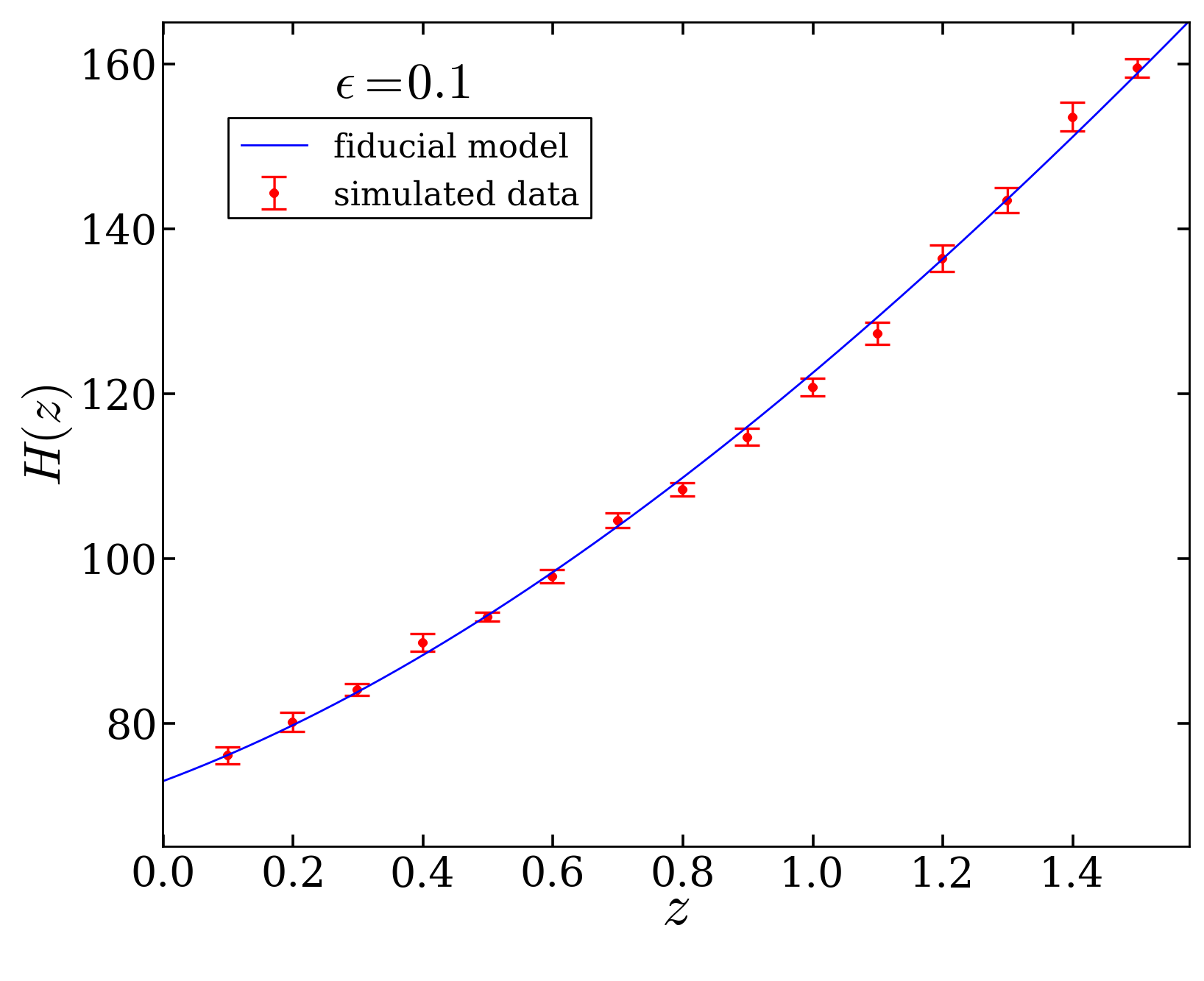}
  	\caption{{\it{Right:}} Similar to Fig. \ref{fig:analytic}, but averaging over the redshift interval [0.1-1.5]. {\it{Left:}} A Monte Carlo realization of 15 data points of the Hubble parameter with 1\% accuracy.}
  	\label{fig:analytic_mean_z}
  \end{figure*}

The result is shown in Fig. \ref{fig:analytic_mean_z} (left) for three different values of $\epsilon$. Note that the relative errors in $\epsilon$ increases considerably with  $\frac{\delta H}{H}$. Therefore, we restrict ourselves to analyze the accuracy required in $H(z)$ measurements to distinguish between a $\epsilon$CDM model and the standard cosmological scenario ($\epsilon = 0$), which is formally determined at the point $\delta \epsilon = \epsilon$ (100\% error). According to these results, we  see that if $\epsilon = 0.1$, an accuracy of $\delta H / H \lesssim 1\%$ would be necessary to rule out $\Lambda$CDM with $1\sigma$ confidence level. For a weaker interaction, however, e.g. $\epsilon = 0.05$, $\delta H / H $ must be smaller than 0.5\%.

\section{Simulations of H(z)}

Current data of the Hubble parameter  has roughly 15\% of uncertainties~\cite{stern}. At such a level, it is not possible to distinguish between the standard $\Lambda$CDM model and a interacting scenario with $\epsilon \lesssim 0.1$. On the other hand, future surveys will be able to measure the expansion rate $H(z)$ with better precision. This is the case, for instance, of the Southern African Large Telescope~\cite{salt} and the Atacama Cosmology Telescope~\cite{act}, which may reach 3\%-10\% accuracy from age estimates of passively evolving galaxies at high-$z$ and large photometric surveys, like J-PAS, that may reach a 2\% level from measurements of the radial scale of the baryonic acoustic oscllations~\cite{surveys}. 

To anticipate the aforementioned project's results concerning a possible interaction in the dark sector, we made use of Monte Carlo simulations to generate samples of $H(z)$ data. The fiducial model in our simulations is a $\epsilon$CDM cosmology, represented by Eq. (\ref{eqH}), with the set of parameters $\mathbf{P} = (H_0, \Omega_{b0}, \Omega_{m0})$ mentioned earlier and some selected values of the interaction parameter $\epsilon$ in the range [-0.2 -- 0.2]\footnote{{Although negative values of $\epsilon$ are forbidden by thermodynamics~\cite{alcaniz-lima}, observational data do not exclude this possibility \cite{ernandes}.}}. For each value of $\epsilon$, we simulated samples with relative error in $H(z)$ of 3\%, 1\%, 0.5\%, 0.3\% and 0.1\%. For each of these combinations, we follow Ref.~\citep{carvalho-alcaniz} and run 500 simulations of 15 points equally spaced {in} redshift corresponding to observations in the range [0.1 -- 1.5].

The simulated data of $H(z)$ are drawn from a normal distribution $N(\mu,\sigma)$ with mean $\mu = H_{fid} $ (see Eq. \ref{eqH}) and standard deviation $\sigma = \left( \frac{\delta H}{H} \right) H_{fid} $, where $\left( \frac{\delta H}{H} \right)$ is the relative error in the determination of $H(z)$.
In order to calculate the error bars we made a bootstrapping study of the current observational errors of $H(z)$ data, as obtained by Stern {\it{et al.}} \cite{stern} using age difference between passively evolving galaxies at different $z$, and estimated the deviation $S$ from the average $15\%$ value. Supposing the errors have a normal distribution, the error bars {are} drawn from the normal distribution
$\sigma_H = N({\sigma}, {S \sigma})$. Fig. 2 (left) shows a Monte Carlo realization of 15 simulated values of the Hubble parameter with 1\% accuracy by assuming $\epsilon = 0.1$.

  \begin{figure*}[htbp] 
 	\centering
 		\includegraphics[scale=0.475]{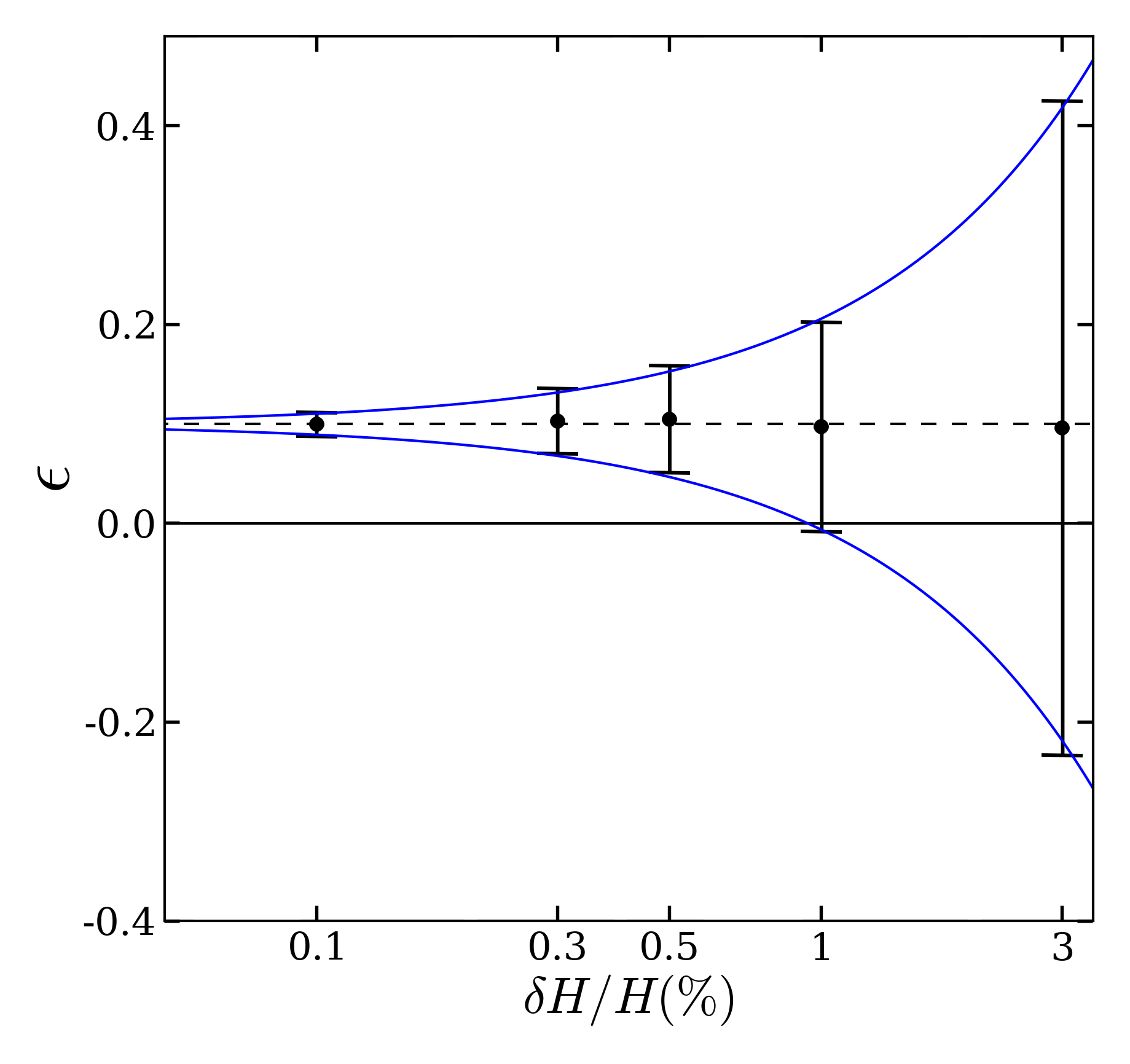}
		\includegraphics[scale=0.475]{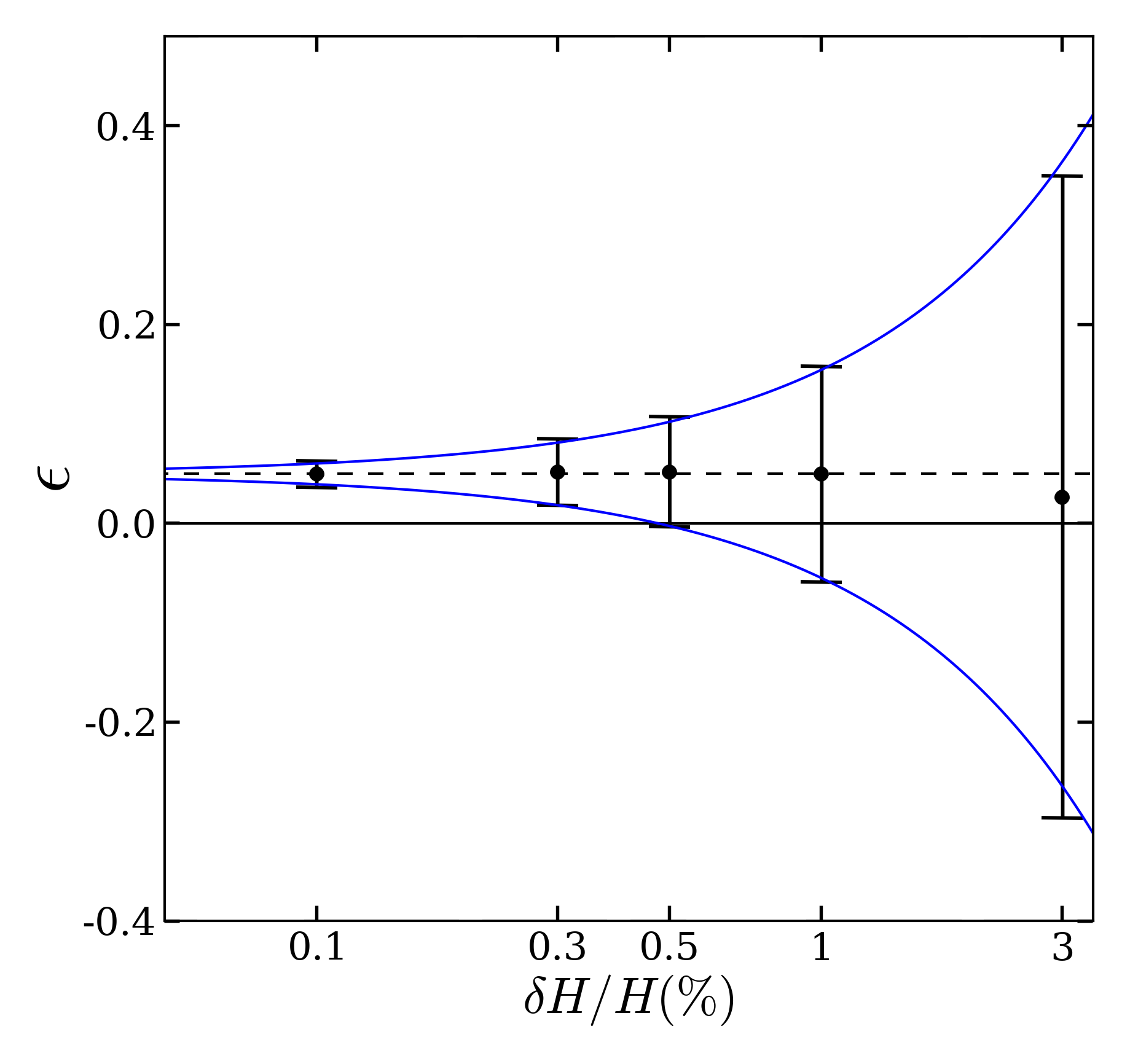}
 	\caption{The average fitted value of $\overline{\epsilon}$ as a function of the relative error in $H(z)$. The left and right panels show, respectively, the fiducial values (dashed line) $\epsilon = 0.1$ and $\epsilon = 0.05$. The blue curves stand for the analytical result derived from Eq. (\ref{eqdeltaH}). }
 	\label{fig:epsilon_01_03}
 \end{figure*}

\section{Results}

Consider a fiducial model characterized by a given value of $\epsilon = \epsilon ' $ and the set of parameters $\mathbf{P}$ given above. For each of the 500 realizations, we minimize the function
\begin{equation}
\chi^ 2 = \sum_i \frac{\left[H(z_i) - H_{fid}(z_i)\right]^2}{\sigma_H^2(z_i)}\;.
\end{equation}
The average $\overline{\epsilon} $ and its standard deviation are determined from the distribution of the resulting 500 best-fit values of $\epsilon$.  Figure \ref{fig:epsilon_01_03} shows the main results of our analyses. Plots of $\overline{\epsilon}$ with error bars corresponding to $1\sigma$ {standard deviation from the mean} are shown for the cases of $\epsilon = 0.1$ (left) and $\epsilon = 0.05$ (right).
From the left panel, we clearly see that if the Hubble parameter reaches a relative error of $\frac{\delta H}{H} < 1\%$, we would be able to distinguish between a $\epsilon$CDM model with $\epsilon = 0.1$ and the standard $\Lambda$CDM scenario at $1\sigma$ level. For $\epsilon = 0.05$ (right panel), a distintion at  $1\sigma$ level between these two classes of models is possible only when the {relative error} of $H(z)$ observation reaches $\frac{\delta H}{H} < 0.5\%$.

For the sake of comparison, we also plot in Fig. \ref{fig:epsilon_01_03} the analytical result derived from Eq. (\ref{eqdeltaH}). Note that it (blue line) shows almost no difference from the simulations, proving to be an accurate tool for cosmological forecasts. It is worth mentioning that we also performed the same analyses presented in Fig. \ref{fig:epsilon_01_03} by considering negative values of $\epsilon$. In this case, we note that this range of values requires a slightly less accuracy in $H(z)$ measurements, with $\frac{\delta H}{H}$ ranging from $\sim 5-10\%$ higher than those corresponding to positive values. The theoretical results from Eq. (\ref{eqdeltaH}) and simulations {also agree for values of $\epsilon < 0$}.

\section{Conclusions}

We performed Monte Carlo simulations of the expansion rate $H(z)$ in a universe with interaction in the dark sector. The analyses have been performed by considering  values of the interacting parameter $\epsilon$ in the interval [-0.2 -- 0.2], which encompasses the current observational bounds on this quantity and have assumed uncertainties on $H(z)$ of $\frac{\delta H}{H} = 3\%$, 1\%, 0.5\%, 0.3\% and 0.1\%. For each combination of $\frac{\delta H}{H}$ and $\epsilon$ we generated 500 Monte Carlo simulations of $H(z)$ in the redshift range [0.1 -- 1.5].

We used these simulations to investigate the possibility of future projects probing the interaction between dark matter and dark energy through the model described by Eq. (\ref{eqH}). We found that values of $\epsilon$ that are within current estimated errors (0.1 and 0.05) might be detected through $H(z)$ observations only if $\delta H/H \lesssim 1\%$. It should be noted that the observational accuracy can be appreciably increased by combining $H(z)$ measurements with data from other observables, such as SNe Ia and CMB data. Currently, these observations weakly constrain $\epsilon$ and in some cases favor negative values \cite{ernandes}.

We also discussed an analytical procedure to foresee $\delta \epsilon / \epsilon$ as a function of $\delta H/H$ [Eq. (\ref{eqdeltaH})] and showed that it provides results in agreement with Monte Carlo simulations. We emphasize that this procedure can be applied to any observable as well as to joint analyses involving different cosmological probes. We plan to extend this analytical approach to combined probes in a forthcoming communication.


\begin{acknowledgments}

The authors acknowledge financial support from CNPq, CAPES and FAPERJ. Part of the results discussed in this paper was carried out with the aid of the Computer System of High Performance of the International Institute of Physics -- UFRN, Natal, Brazil.

\end{acknowledgments}

\end{document}